\documentstyle[a4,12pt,epsf]{article}
\begin{document}
\begin{center}
{\large\bf Soft pions at high energy as an origin of flavor asymmetry
of the light sea quarks in the nucleon}\\
Susumu  Koretune \\
Department of Physics,Shimane Medical University,\\
Izumo,Shimane,693-8501
\end{center}
By using the soft pion theorem in the inclusive reactions, soft pions' contribution
to the structure function $F_2$ in the nucleon is estimated. It is shown that
this contribution produces such a large flavor asymmetry in the light sea
quark distributions that it gives about $30\sim 50 \%$ of the NMC deficit in the
Gottfried sum.

\section{Introduction}
The modified Gottfried sum rule \cite{Got} has explained the NMC deficit in the
Gottfried sum \cite{NMC} almost model independently. It has shown that
the deficit is the reflection of the hadronic vacuum originating from the
spontaneous chiral symmetry breakings. In this sense the physics underlining
this algebraic approach has a common feature with that of the mesonic 
models reviewed in \cite{meso}. However,in the algebraic approach,
importance of the high energy region not only in the theoretical
meaning but also in the numerical analysis has been made clear.
Further the numerical prediction based on this
sum rule exactly agrees with the recent experimental value from E866/NuSea
collaboration \cite{E866}. This experiment also gives us the light antiquark
difference $(\bar{d}(x)-\bar{u}(x))$ and the ratio $\bar{d}(x)/\bar{u}(x)$
in the range $0.02\leq x \leq 0.345$. An unexpected behavior is that the
asymmetry seems to dissapear at large $x$. On the other hand, a typical
calculation in the mesonic models based on the $\pi NN$ and the $\pi N\Delta$ 
processes account for about a half of the NMC deficit \cite{meso}. 
According to the E866 experiment, an explanation of the remaining half of 
the NMC deficit should be given by contributions in the medium or 
the small $x$ region. Unfortunately, the approach from the mesonic models
can not account for the magnitude from these regions definitely.
In fact, the $\pi N\Delta$ process partly cancels the positive contribution
to the $(\bar{d}(x)-\bar{u}(x))$ from the $\pi NN$ process. The contributions
from the higher resonances or from the multiparticle states are obscure.
Hence the best we can say is that the mesonic models explain the flavor
asymmetry of the light sea quarks qualitatively. These facts suggest
that there may exist a dynamical mechanism so far overlooked to produce
the flavor asymmetry at medium and high energy, and that it may compensate
the above flaw of the mesonic models. In this paper, it is shown that
the soft pion theorem in the inclusive reaction at high energy \cite{sakai}can
explain about $30\sim 50\%$ of the NMC deficit where we take the magnitude of it
as $0.07$ following the E866 experiment\cite{E866} for definiteness.
\section{Soft pions at high energy}
Since the soft pion theorem in the inclusive reaction at high energy
is not well known, let us first explain it briefly. Usually,the soft
pion theorem has been considered to be applicable only in the low
energy regions. However in \cite{sakai}, it has been found that this theorem
can be used in the inclusive reactions at high energy if the Feynman's
scaling hypothesis holds. In the inclusive reaction ``$\pi + p \to
\pi_{s}(k) + anything$'' with the $\pi_{s}$ being the soft pion, 
it states that the differential cross-section
in the center of the mass (CM) frame defined as
\begin{equation}
f(k^3,\vec{k}^{\bot},p^0)=k^0\frac{d\sigma}{d^3k} ,
\end{equation}
where $p^0$ is the CM frame energy, scales as
\begin{equation}
f\sim f^{F}(\frac{k^3}{p^0},\vec{k}^{\bot}) + \frac{g(k^3,\vec{k}^{\bot})}{p^0} .
\end{equation} 
If $g(k^3,\vec{k}^{\bot})$ is not singular at $k^3=0$, we obtain
\begin{equation}
\lim_{p^0\to \infty}f^F(\frac{k^3}{p^0},\vec{k}^{\bot}=0)=
f^F(0,0)=\lim_{p^0\to \infty}f(0,0,p^0) .
\end{equation}
This means that the $\pi$ mesons with the momenta $k^3<O(p^0)$ and
$\vec{k}^{\bot}=0$ in the CM frame can be interpreted as the soft pion.
This fact holds even when the scaling violation effect exists,
since we can replace the exact scaling by the approximate one in this
discussion. In Weinberg's language, these soft pions correspond to
semi-soft pions \cite{wein}. The important point of this soft pion 
theorem is that the soft-pion limit can not be interchanged with the
manipulation to take the 
discontinuity of the reaction ``$a + b + \bar{\pi_s} \to a + b + \bar{\pi_s}$''. 
We must first take the soft pion limit in the reaction ``$a + b \to \pi_s + anything$''.
This is because the soft pion attached to the nucleon(anti-nucleon)
in the final state is missed if we take the discontinuity of the soft pion limit of
the reaction
``$a + b + \bar{\pi_s} \to a + b + \bar{\pi_s}$'' \cite{sakai}.

Now, based on the null-plane formalism, this soft pion theorem has been extensively 
studied in \cite{kore78}. In the usual equal-time formalism,
the contribution where the soft pion attached to the nucleon(anti-nucleon)
depends on its velocity because the one particle helicity matrix element
of the axial vector current takes the form
$<p,h|J_a^{50}(0)|p,h^{\prime}> = 2hp^0g_{A}(0)v\delta_{hh^{\prime}}$,
where $h,h^{\prime}$ denote the helicity and
$v=|\vec{p}|/p^0$. On the other hand in the null-plane formalism it takes the form
$<p,h|J_a^{5+}(0)|p,h^{\prime}> = 2hp^+g_{A}(0)\delta_{hh^{\prime}}$, where
the light-like helicity base is used in this case. 
Hence in the null-plane formalism the velocity factor is always 1, and the
ambiguity from this part disappears. By using light-cone current 
algebra \cite{fg} in the inclusive lepton-hadron scatterings,
theoretical prediction in the case of the soft $\pi^-$ was compared
with the data of the $\pi^-$ production in the central region and it
has been suggested that the mechanism proposed in \cite{sakai} should
be applied to the directly produced pions,i.e.,the pions produced
not through the decay from the resonance\cite{kore78reso}.
In \cite{kore82}, the cut vertex formalism \cite{Mue} has been used
in stead of the light-cone current algebra, and the charge asymmetry
in the central region in the inclusive lepton-hadron scatterings is
considered. This is because the pions from the resonance decay product 
due to the strong interaction cancels out in the asymmetry in the
central region, and hence the experimentally measured asymmetry is mainly 
due to the directly produced pions. It has been found that the experimental 
value roughly agrees with the theoretical expectation based on the
soft pion theorem in the inclusive reactions. Several years ago,
the photoproduction version of the modified Gottfried sum rule
has been studied, and found that the soft pions' contribution
at high energy plays an important role to satisfy the
sum rules \cite{kore93}.
\section{Contribution to the Gottfried sum}
Let us now consider the reaction 
``$\gamma_{V}(q)$ + nucleon(p) $\to \pi_{s}(k)$
+ anythings'', where $\gamma_{V}$ means the virtual photon.
We take the soft pion limit $k^{\mu}\to 0$ by first taking
$\vec{k}^{\bot}=0,k^+=0$ and then $k^{-}\to 0$ in the
scattering amplitude. In this limit we can classify it
into three kinds of terms. Type(a) term is the amplitude in which the proper
part of the axial-vector current attaches to the initial nucleon.
Type(b) term is the amplitude in which the proper part of the axial-vector
current attaches to the final nucleon or anti-nucleon. Type(c) term
is the amplitude which comes from the commutation relation on the 
null-plane. Then by taking the square of the amplitude in the
soft pion limit, we construct the hadronic tensor.
Following \cite{kore78}, we classify the contribution
to the hadronic tensor as follows: The term coming from the
type $a^{\dagger}a$ is $A_1^{\mu \nu}$, $a^{\dagger}c+c^{\dagger}a$
is $A_2^{\mu \nu}+A_3^{\mu \nu}$, $b^{\dagger}b$ is $B_1^{\mu \nu}$,
$b^{\dagger}c+c^{\dagger}b$ is $B_2^{\mu \nu}+B_3^{\mu \nu}$,
$a^{\dagger}b+b^{\dagger}a$ is $C_1^{\mu \nu}$, and $c^{\dagger}c$
is $D_4^{\mu \nu}$, where $a,b,$ and $c$ denotes the type of the
amplitude in the soft pion limit. 
Now in the inclusive reactions the kinematic variables in the 
initial state are unconstrained in the soft pion limit.
We can take the usual deep-inelastic limit.
The hadronic tensor is light-cone dominated in the
deep-inelastic limit,hence we can use the light-cone
current algebra, and find how the soft pion piece
is related to the structure functions in the total inclusive
reactions. In the perturbative analysis in QCD the $Q^2$ dependence 
can be taken into account by the cut vertex formalism suitable for the
light-cone dominated processes. This is because in our case the hadronic tensor 
is not short distance dominated in the short distance limit as in the hadonic 
tensor in the total inclusive reaction where it is expressed by the matrix
element of the commutation relation of the currents. 
In the soft pion limit surviving pole terms
are restricted by the pion's charge. For example in the $\pi_s^-$
case, proper part of the axial-vector current attached to 
the initial proton is prohibited by the charge conservation.
Because of the asymmetry of this kind we encounter the terms
which can not be expressed by the commutation relation,
which prevents us to show the short distance dominance
in the short distance limit. Thus the usual method which makes the short
distance expansion first and then continue it analytically to the light-cone
with use of the dispersion relation can not be applied. The asymmetry 
discussed above together with
the fact that nucleon charge is changed when the
proper part of the axial-vector current corresponding to
the charged pion is attached to the nucleon is the origin
of the charge asymmetry in the soft pion limit.
Now the contribution from $B_2^{\mu \nu}+B_3^{\mu \nu}$ and
$C_1^{\mu \nu}$ can be neglected in the deep-inelastic region.
In these terms, the positive helicity state of the final nucleon
(anti-nucleon) and the negative one contribute oppositely 
in sign, hence their contribution at high energy can be expected
to be very small, while the contribution in the low energy
region is suppressed by the form factor effect in the deep-inelastic
region. Thus we consider the contribution only from 
$A_1^{\mu \nu}$, $A_2^{\mu \nu}+A_3^{\mu \nu}$, $B_1^{\mu \nu}$,
and $D_4^{\mu \nu}$. Since the detailed expressions are given
in \cite{kore78,kore82}, it is straightforward to obtain
the soft pions' contribution to the structure function
$F_2$. Adding the contributions from the soft $\pi_s^+,
\pi_s^-,$ and $\pi_s^0$, and subtracting the contributions to
$F_2^{en}$ from those to $F_2^{ep}$, we obtain
\begin{eqnarray}
\lefteqn{(F_2^{ep} - F_2^{en})|_{soft}}\nonumber \\
 &=& \frac{I_{\pi}}{4f_{\pi}^2}[g_A^2(0)(F_2^{ep} - F_2^{en})(3<n> -1)
-16xg_A(0)(g_1^{ep} - g_1^{en})] ,
\end{eqnarray}
where $I_{\pi}$ is the phase space factor for the soft pion defined as
\begin{equation}
I_{\pi} = \int\frac{d^2\vec{k}^{\bot}dk^+}{(2\pi)^{3}2k^+}
\end{equation}
where $<n>$
is the sum of the nucleon and anti-nucleon multiplicity defined as
$<n>=<n>_p + <n>_n + <n>_{\bar{p}} + <n>_{\bar{n}}$.
In Eq.(4), the contribution coming from $D_4^{\mu \nu}$ cancels
out, among the terms proportional to $g_A^2(0)$ the one which has 
a factor $<n>$ comes from $B_1^{\mu \nu}$ and the other one
comes from $A_1^{\mu \nu}$, and the term proportional to
the spin dependent function $(g_1^{ep} - g_1^{en})$ comes from
$A_2^{\mu \nu}+A_3^{\mu \nu}$. Note that this spin dependent term
is obtained in the approximation to neglect the sea quarks' contribution to $(g_1^{ep} - g_1^{en})$. 
Without this approximation $16(g_1^{ep} - g_1^{en})$ in Eq.(4) should be replaced by
$\displaystyle{24(g_1^{ep} - g_1^{en})-\frac{4}{3}(g_1^{\bar{\nu} p} - g_1^{\nu p})}$.

Now as explained, the soft pions' contribution in the inclusive reaction
can not be obtained by the discontinuity formula in the sense that the interchange
of taking the soft pion limit and taking the discontinuity is impossible.
Because of this fact, we must revise the structure function 
$(F_2^{ep} - F_2^{en})$ as 
$(F_2^{ep} - F_2^{en})=(F_2^{ep} - F_2^{en})_u +
(F_2^{ep} - F_2^{en})|_{soft}$, where the suffix $u$
specifys the usual one which satisfys the generalized unitarity. 
In the parton model, using impulse approximation,
structure function is obtained as an imaginary part of the incoherent
elastic scattering of the virtual photon off quarks. Thus the soft pion
piece is not included in the parton model in general. However in the
deep inelastic region we parametrize the structure function by the quark
distribution functions. Hence we should revise them to include the soft pions'
contributions.
Now the soft pion contributes to the structure function $(F_2^{\nu p}-F_2^{\bar{\nu} p})$
and the Adler sum rule fixes the valence quark distribution as
$\int_0^1dx(u_v - d_v)=1$. Hence the phenomenologically determined valence
quark distribution $(u_v -d_v)$ which satisfys the constraint
already effectively takes the contribution from the soft pion piece since
the Adler sum rule is satisfied only if this contribution is taken into account.
Then we use these valence quark distributions to fit the structure function
$(F_2^{ep} - F_2^{en})$. Therefore the soft pion piece $(F_2^{ep} - F_2^{en})|_{soft}$
should be effectively taken into account in the phenomenologically 
determined sea quark distributions.    
Thus, by assuming the light sea quark distribution being equal
to its antiquark distribution for simplicity,
we can express $(F_2^{ep} - F_2^{en})|_{soft}$ as the asymmetry of the antiquark 
distribution as
\begin{equation}
(F_2^{ep} - F_2^{en})|_{soft} = -\frac{2}{3}x(\bar{d} - \bar{u})|_{soft}.
\end{equation}
To estimate the magnitude of this asymmetry, we approximate $F_2^{ep},
F_2^{en},g_1^{ep},g_1^{en}$ on the right-hand side of Eq.(4) by the
valence quarks distribution functions at $Q_0^2=4\; GeV^2$\cite{GS}.
As a multiplicity of the nucleon and antinucleon, we set
\begin{equation}
<n> = a\log_es +1 ,
\end{equation}
where $s=(p+q)^2$. The parameter $a$ is fixed as
0.2 in consideration for the proton and the anti-proton multiplicity
in the $e^+e^-$ annihilation such that $\frac{1}{2}a\log_es$ with 
$\sqrt{s}$ replaced by CM energy of that reaction agrees 
with it \cite{DELPHI}. 
As to the pion phase space factor $I_{\pi}$, we estimate it as follows.
We assume approximate Feynman scaling and regard the directly produced
pions in the virtual-photon and the target-nucleon center of the mass (CM)
frame which satisfy the two conditions as soft pions. 
\begin{description}
\item[(1)]The transverse momentum satisfys $|\vec{k}^{\bot}|\leq bm_{\pi}$.
\item[(2)]Feynman scaling variable $x_{F}=2k^3/\sqrt{s}$ satisfys
$|x_{F}|\leq c$.
\end{description}
\begin{figure}
\epsfbox{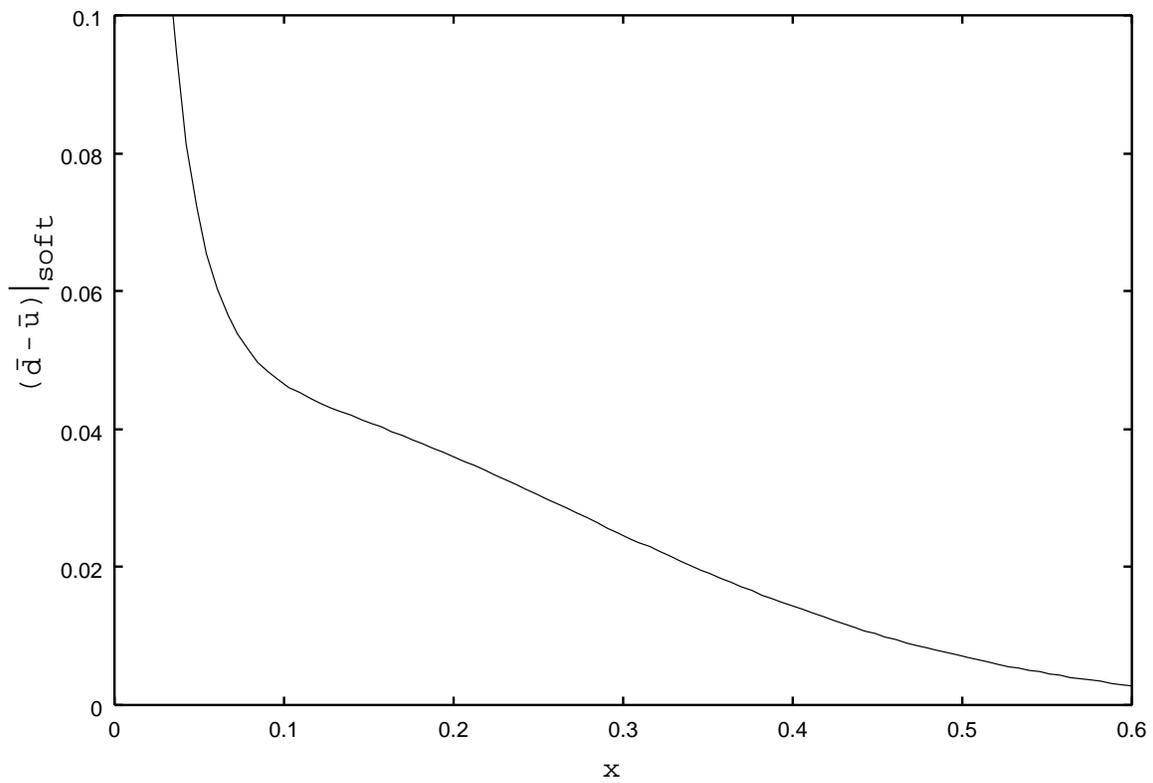}
\caption{soft pions' contribution to $\bar{d}-\bar{u}$}
\label{fig:1}
\end{figure}
Here we take the momentum $k$ in the CM frame, and consider the constant $b$ 
as the value near 1, and $c$ near 0.1. These values are fixed based on the
previous works\cite{kore78reso,kore82} which showed that the directly
produced pions in the central region in the CM frame expected by the 
experimentally measured quantity were the same order
with the soft pions' contribution.  The experimentally expected values
were always larger but the difference were within factor 2. 
The upper and the lower limit of the integral with respect to $k^+$ in 
the phase space factor $I_{\pi}$ is restricted by the condition (2).
The lower limit of the $k^+$ behaves as $O(1/\sqrt{s})$ at
high energy. Because the soft pion limit is the finite part as $k^{\mu}\to 0$,
the factor $1/k^+$ in $I_{\pi}$ greatly enhance the soft pions' contribution.
By doing the explicit integration we obtain the phase space factor $I_{\pi}$ as
\begin{eqnarray}
I_{\pi}&=&\frac{1}{16\pi^2}\Bigg( (b^2+1)m_{\pi}^2\log_e \left(
\frac{\sqrt{(1+b^2)m_{\pi}^2+\frac{c^2s}{4}}+\frac{c\sqrt{s}}{2}}
{\sqrt{(1+b^2)m_{\pi}^2+\frac{c^2s}{4}}-\frac{c\sqrt{s}}{2}}\right)\\\nonumber
&-&m_{\pi}^2\log_e \left( \frac{\sqrt{m_{\pi}^2+\frac{c^2s}{4}}+\frac{c\sqrt{s}}{2}}
{\sqrt{m_{\pi}^2+\frac{c^2s}{4}}-\frac{c\sqrt{s}}{2}}\right)
+c\sqrt{s}\left( \sqrt{(1+b^2)m_{\pi}^2+\frac{c^2s}{4}}-
\sqrt{m_{\pi}^2+\frac{c^2s}{4}}\right) \Bigg)   .
\end{eqnarray}
A typical example of the antiquark asymmetry $(\bar{d} - \bar{u})|_{soft}$
given by Eqs.(4) and (6) is given in the fig.1 for $a=0.2,b=1,c=0.1$
in the range $0.05\leq x \leq 0.6$. 
Further to grasp the soft pions' contribution to the asymmetry $(\bar{d} - \bar{u})$ 
qualitatively, we plot the value of $x(\bar{d} - \bar{u})$ of
the CTEQ4M\cite{CTEQ} fit at $Q^2=4\; GeV^2$ in the fig.2.
\begin{figure}
\epsfbox{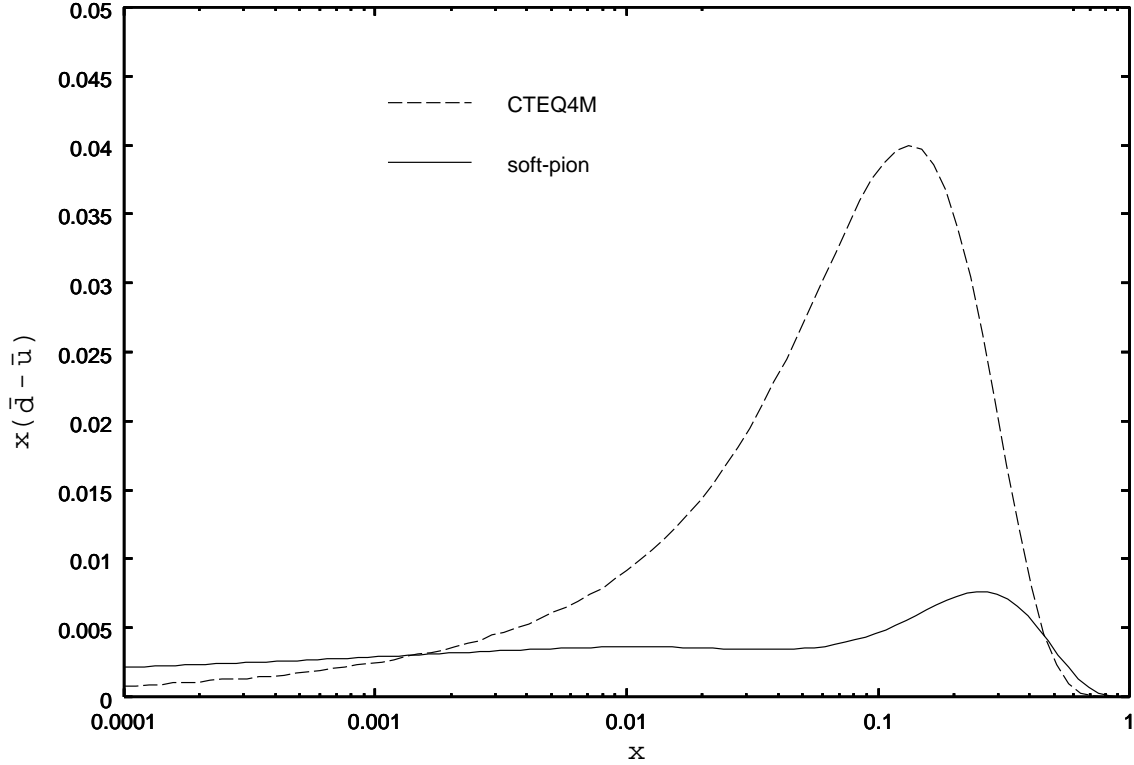}
\caption{soft pions' contribution to $x(\bar{d}-\bar{u})$}
\label{fig:2}
\end{figure}
From the fig.2 we can recognize that the soft pions' contribution to the
Gottfried sum is large because the small $x$ tail is slowly decreasing.
However extrapolation of the theoretical curve to the very small $x$ region
can not be trusted because the input distribution can not be trusted
in the very small $x$ region. Hence we should cut the integral somewhere
in the very small $x$ region. While, theoretical expectation of
the contribution above $x=0.3$ may be large, but the contribution
from this region to the Gottfried sum is small. Further the phase space constraint
from this region may become more stringent. In any case it rapidly
becomes zero as we go to large $x$.  Now in the medium $x$ region, 
the small bump in the the fig.2 may be related to the small excess of the E866 data 
compared with the contribution from the meson cloud model\cite{E866} 
since the soft pions' contribution 
should be added to the contribution to the meson cloud model as a background contribution.
By taking these fact into consideration, we investigate
$J(\alpha,\beta )=\int_{\alpha}^{\beta}\frac{dx}{x}(F_2^{ep}-F_2^{en})|_{soft}$
for various values of $a,b,c$. For $a=0.18,b=1,$ and $c=0.1$,$J(10^{-4},0.2)=-0.019,
J(10^{-5},0.2)=-0.022,J(10^{-6},0.2)=-0.024$. For $a=0.20,b=1,$ and $c=0.1$,
$J(10^{-4},0.2)=-0.018,J(10^{-5},0.2)=-0.021,J(10^{-6},0.2)=-0.023$.
For $a=0.22,b=1,$ and $c=0.1$,$J(10^{-4},0.2)=-0.017,J(10^{-5},0.2)=-0.019,
J(10^{-6},0.2)=-0.021$. Thus the effect of the change of $a$ which is consistent
to the experimental value of the multiplicity data of the $e^+e^-$
experiment is small. The extrapolation of the integral to the  
smaller value of $x$ make the value of $J$  smaller, but its magnitude
is not so large. For example for $a=0.20,b=1,$ and $c=0.1$,$J(10^{-9},1)=-0.030$
and $J(0.2,1)=-0.005$. As the effect of the change of $c$, we take the case
$a=0.2, b=1,$ and $c=0.05$ and obtain $J(10^{-4},0.2)=-0.012, J(10^{-5},0.2)=-0.015,
J(10^{-6},0.2)=-0.017$. Thus the effect of this change is $25\%$ reduction
compared with the case $c=0.1$. We consider that $b$ takes the value
near 1 except in the large $x$ region at low energy, where
the allowed phase space becomes the ball rather than the cylinder defined 
by the condition (1) and (2). This causes the more rapid decrease at large $x$ than 
the one given in fig.2.  However the change in this region does not give a sizable
effect to the value of $J$. 
Though we can not say the exact magnitude, we see that soft pions'
contribution gives the sizable effect to the 
NMC defect. Based on the above analysis we estimate 
that $J(0,1)$ takes the value about $-0.04 \sim -0.02.$
\section{Conclusion}
The soft pion theorem in the inclusive reaction is very general
and is useful if the approximate scaling such as the Feynman scaling holds.
The magnitude of the contribution is non-negligible as is shown 
in \cite{kore78reso,kore82} and also in this example.
In fact it can reach about $30\sim 50\%$ of the NMC deficit.
This is just the one lacked in the typical calculation
in the mesonic models\cite{meso}. The main contribution from the soft pion comes 
from the medium and the high energy regions where the mesonic model lacks
its predictive ability and where the algebraic approach has pointed out
its importance.

\end{document}